# Gate modulation of the long-range magnetic order in a vanadium-doped WSe$_2$ semiconductor

Dinh Loc Duong[1,2,*], Seong-Gon Kim[3], Young Hee Lee[1,2,4,*]

[1]Center for Integrated Nanostructure Physics (CINAP), Institute for Basic Science (IBS), Suwon 16419, Republic of Korea.

[2]Department of Energy Science, Sungkyunkwan University, Suwon 16419, Republic of Korea.

[3] Department of Physics and Astronomy, Mississippi State University, Mississippi State, MS 39762, USA

[4]Department of Physics, Sungkyunkwan University, Suwon 16419, Republic of Korea.

We demonstrate the gate-tunability of the long-range magnetic order in a p-type V-doped WSe$_2$ monolayer using *ab initio* calculations. We found that at a low V-doping concentration limit, the long-range ferromagnetic order is enhanced by increasing the hole density. In contrast, the short-range antiferromagnetic order is manifested at a high electron density by full compensation of the p-type V doping concentration. The hole-mediated long-range magnetic exchange is ~70 meV, thus strongly suggesting the ferromagnetism in V-doped WSe$_2$ at room temperature. Our findings on strong coupling between charge and spin order in V-doped WSe$_2$ provide plenty of room for multifunctional gate-tunable spintronics.

Magnetic semiconductors, which reveal the electrical manipulation of magnetic properties, are expected to be candidates for low-power spintronic devices [1]. Although the antiferromagnetic semiconductors and insulators are promising for high-speed spin transport devices [2–9], the existence of room-temperature ferromagnetic semiconductors remains controversial, for either intrinsic or diluted ferromagnetic semiconductors [10–13]. Recently, room-temperature ferromagnetic domains in V-doped $WSe_2$ have been observed [14]. The proposed mechanism of the long-range ferromagnetic order is Ruderman–Kittel–Kasuya–Yosida or Zener interaction, where the free holes play a role as a medium to establish the interaction between V atoms [15,16].

One primary feature of carrier-mediated ferromagnetic semiconductors is their capability of manipulating the magnetic properties or spin states through electrons by modulating the carrier density or external electric field [17–20]. This allows for multifunctional spintronic devices to be realized [1,21]. Here, we investigate the response of the magnetic properties of V-doped $WSe_2$ with different electron-hole densities introduced by carrier injection using density functional calculations. We found that the long-range ferromagnetic order can be modulated by changing the carrier density of the V-doped $WSe_2$. Furthermore, fully compensated electron doping can completely turn the ferromagnetic ground state into a short-range antiferromagnetic order.

The band structure of V-doped $WSe_2$ with spin-orbit coupling was calculated using the Quantum Espresso code [22]. Projector-augmented wave potentials were used with a cut-off energy of 30 Ry [23], the convergence test for which was confirmed by considering the band order [16]. To describe the strong correlation of the $d$ electron, GGA+U method with U=3 was used for the vanadium atom. The initial spin state is induced along z direction. Validation of the band structure calculation was confirmed in our previous study using scanning tunneling spectroscopy and a field-effect transistor [15].

Figure 1a depicts the band structure of pure WSe$_2$ including spin-orbit coupling (SOC), which reveals a band gap of 1.3 eV with the Fermi level located in the middle of the band gap. The density of states (DOS) corresponding to the magnetic quantum number $m_{j=5/2}$ = -1/2 (red) and $m_{j=5/2}$ = +1/2 (blue) (the right panel of Fig. 1a) manifests an equal number of DOS between two magnetic quantum states, implying the non-magnetic characteristics of pristine WSe$_2$. By contrast, the V-doped WSe$_2$ band structure (Fig. 1b) manifests the Fermi level shifted to inside the valence band, ensuring the p-type doping effect in vanadium [15,16]. In addition, a discrepancy in the DOS between $m_{j=5/2}$ = -1/2 and $m_{j=5/2}$ = +1/2 in the W site is clearly observed, thereby indicating the presence of spin-polarized states. Correspondingly, the spin-polarized states are clearly revealed in the DOS of the V site with a high magnitude (right panel of Fig. 1b). The total magnetization of the W atoms far from the V site is ~0.003 $\mu_B$, which is lower than ~1.077 $\mu_B$ of the V atom [16].

To investigate the possibility of the gate-tunable magnetic properties of V-doped WSe$_2$, the charge is injected into the system. Figure 1c shows the total DOS of V-doped WSe$_2$ with different injected carrier densities. Because of the p-type doping effect of the V atom, free holes are clearly revealed on the top of the valence band in the V-doped WSe$_2$ monolayer. The Fermi level is shifted deeper inside the valence band with the hole injection. By contrast, the Fermi level is shifted toward the conduction band edge with electron injection. Interestingly, the Fermi level is shifted to the middle of the band gap in V-doped WSe$_2$ when a single electron per 8x8 supercell is injected, indicating that the acceptor state of the V atom is completely filled. In addition to the shifting of Fermi level, the band structure is also significantly modified with electron injection, whereas it remains nearly unchanged with hole injection. We later discuss the band structure of electron injection.

Since the long-range ordering in V-doped $WSe_2$ is mediated by free holes [15,16], investigating the changes in magnetic properties with different carrier densities is intriguing. Figure 2 shows the exchange energy, which is defined as the energy difference between the magnetic and non-magnetic states, and the corresponding total magnetic moments of V-doped $WSe_2$ with different injected carrier densities. The exchange energy becomes stronger when holes are injected and the ferromagnetization accumulated near V atoms is enhanced. This implies that the magnetic moment is modulated with carrier density, which is concrete evidence of the hole-mediated long-range spin interaction induced by the hybridization between the impurity levels of V atoms and valence band edge of $WSe_2$ [15,16]. At a very high hole density, hole carriers are screened, thereby reducing the total ferromagnetization. However, the exchange energy becomes weaker with electron injection, and the magnetic moment approaches nearly zero at a higher electron density, which manifests the tendency of an antiferromagnetic state. Indeed, antiferromagnetism emerges at the charge compensation point, at which the number of injected electrons (-1e) is equal to that of intrinsic holes induced by V atoms.

To confirm the presence of the antiferromagnetic state, the magnetic moments located at each atom are calculated (Table 1). The magnetic moments of W sites near the V atoms have the opposite sign (spin-down) of that at the V site (spin-up) owing to their strong local exchange interaction, which is distinct from all spin-up (ferromagnetism) configuration in the high hole injection state (+2e). At the charge compensation point, the ferromagnetic moment (spin-up) of the W atoms far from V are diminished. Meanwhile, the antiferromagnetic moment (spin-down) of the W atoms near the V sites is enhanced. Furthermore, the spin-up magnetic moment of the V atoms is reduced, thereby facilitating the formation of the antiferromagnetic state. The magnetic moment located at the Se sites is more narrowly localized than that at W sites. There is no magnetic moment located at Se atoms farther than 7 Å from the V site.

It is intriguing to estimate the strength of the interaction energy mediated by free holes. The interaction energy involves two components: i) short-range interaction energy produced by strong hybridization in d-p-d orbitals of V-Se-W atoms (e.g., direct, super, or double exchanges), and ii) long-range interaction energy via holes. The hole-mediated long-range interaction energy can be estimated by the exchange energy difference of the systems with and without holes (as indicated by the arrow of Fig. 2), which is approximately 70 meV. This value is much larger than 25 meV of the thermal energy at 300 K, thus explaining the existence of the long-range ferromagnetic order in V-doped $WSe_2$ at room temperature [14].

The spin density in real space is calculated by injecting one hole (Fig. 3a) and one electron in the supercell (Fig. 3b). The isosurface of 0.0001 $\mu_B/Å^3$ of the spin density (top panels) and projected total magnetic moments along the z direction (bottom panels) are shown correspondingly. Large ferromagnetic moments are clearly manifested at the V and W sites near the V atom. The long-range magnetic order is still maintained with hole injection, implying the existence of a ferromagnetic moment on the W atoms far from the V site (blue spots in Fig. 3a). In contrast, the ferromagnetic moment at the W atoms far from the V site is completely suppressed for one electron injection, whereas high magnetic moments are still present on the V and W/Se atoms near the V site. The magnetic state is antiferromagnetic with a zero net magnetic moment of the supercell. The net magnetic moment of the W atoms far from the V site disappears, as shown in Fig. 3b, indicating a local antiferromagnetic state by the V-Se-W complex without free holes.

Figure 4 depicts the antiferromagnetic band structure of V-doped $WSe_2$ doped with one electron. The number of bands decreases compared to the ferromagnetic states as an inherent antiferromagnetic state. In addition, the empty doping state near the conduction band edge, which is present in the ferromagnetic state, disappears in the antiferromagnetic band structure. The projected DOS of the d orbital of the V atom is shown in the second panel. The contribution of the d orbital in the V atom to the strong hybridization states near the valence band is retained, which is confirmed by the large DOS of V at that energy range. The spin polarization in the V atom is conserved with a smaller absolute magnetic moment as compared to when no carrier injection is conducted (see Table 1). Interestingly, the DOS of W atoms far from the V site (last panel) is still spin-polarized despite a zero net magnetic moment. This implies the presence of a long-range d-d hybridization between the V and W atoms even without free holes in the system. This long-rang interaction is completely different from the carrier-mediated Zener type (kinetic p-d exchange). Instead, it is rather similar to the secondary super-exchange interaction proposed by Katayama-Yoshida model [24,25]. However, the method to distinguish between the degree of long-range order between the d-d and p-d hybridizations remains unresolved and should be addressed in a future study.

We note that the required electron density to fully compensate holes for 8 x 8 V-doped $WSe_2$ (1.6 at.% of V) is approximately $1.6 \times 10^{13}$ $cm^{-2}$, which can be fully compensated by the normal $SiO_2$ or high-k oxide gates. Furthermore, the V-doping concentration can be reduced to this critical electron density, which is validated by our calculation at a V concentration of 1 at.% (10 x 10 supercell).

In conclusion, the long-range ferromagnetic order in V-doped $WSe_2$ is modulated by carrier injection. Charge injection can modulate both the exchange energy and net magnetic moment. More interestingly, the transition to the antiferromagnetic from the ferromagnetic state is possible when carriers are fully compensated. Our prediction demonstrates the gate-tunable magnetic properties of diluted magnetic 2D semiconductors, thus providing a better understanding of the V-doped $WSe_2$ system.

**ACKNOWLEDGEMENTS**

This work was supported by the Institute for Basic Science of Korea (IBS-R011-D1) and the Creative Materials Discovery Program through the National Research Foundation of Korea (NRF) funded by the Ministry of Science, ICT, and Future Planning (No. 2016M3D1A1919181).

**REFERENCES**

* ddloc@skku.edu, leeyoung@skku.edu

**FIGURES AND FIGURE CAPTIONS**

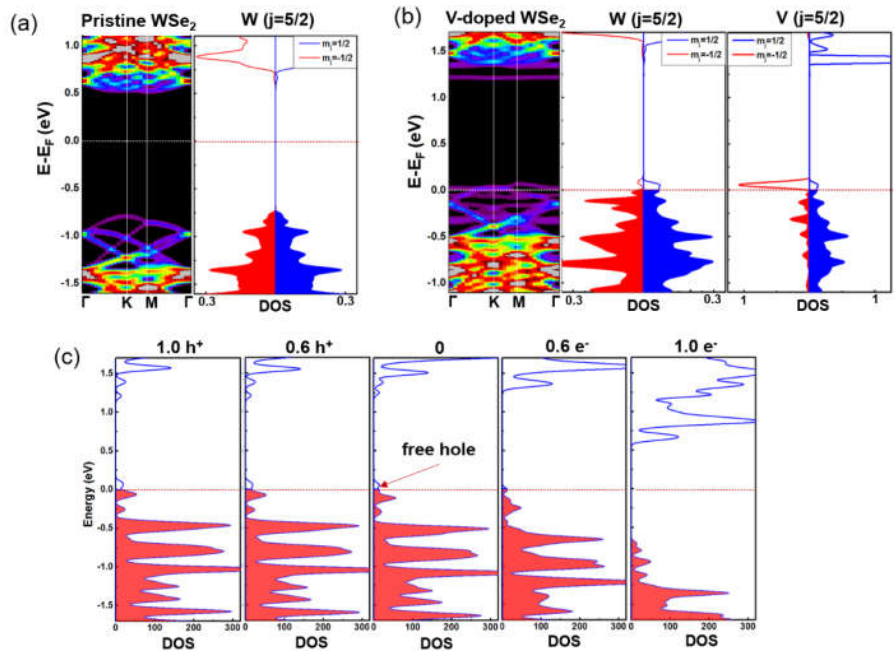

FIG. 1. Spin-polarized band structure and projected DOS located at W atoms of (a) pristine and (b) V-doped WSe$_2$ with SOC. In V-doped WSe$_2$, the projected DOS of W atoms far from the V site and V atom are shown (middle and right panels of (b)). (c) Total DOS of V-doped WSe$_2$ with different doping carrier densities.

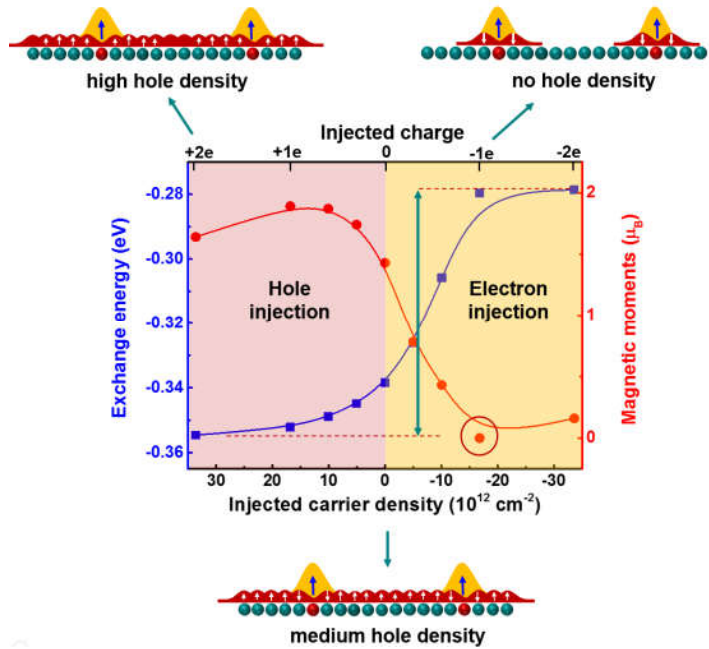

FIG. 2. Exchange energy and net magnetic moments of V-doped WSe$_2$ with different carrier doping densities, and the corresponding schematic models for the magnetic interaction through free holes. The carrier compensation point is indicated by the red circle.

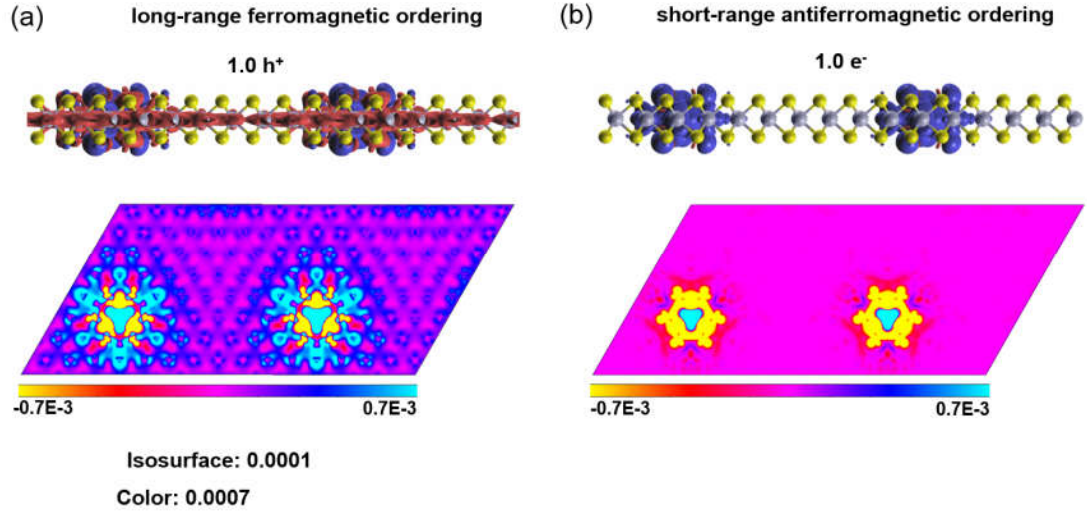

FIG. 3. Isosurface (top) and projected magnetic moments along the c axis (bottom) in the cases of injecting (a) one hole and (b) one electron. This isosurface value is 0.0001 $\mu_B/Å^3$. The color map range is from -0.0007 to 0.0007 $\mu_B/Å^2$ for spin-down and spin-up states, respectively.

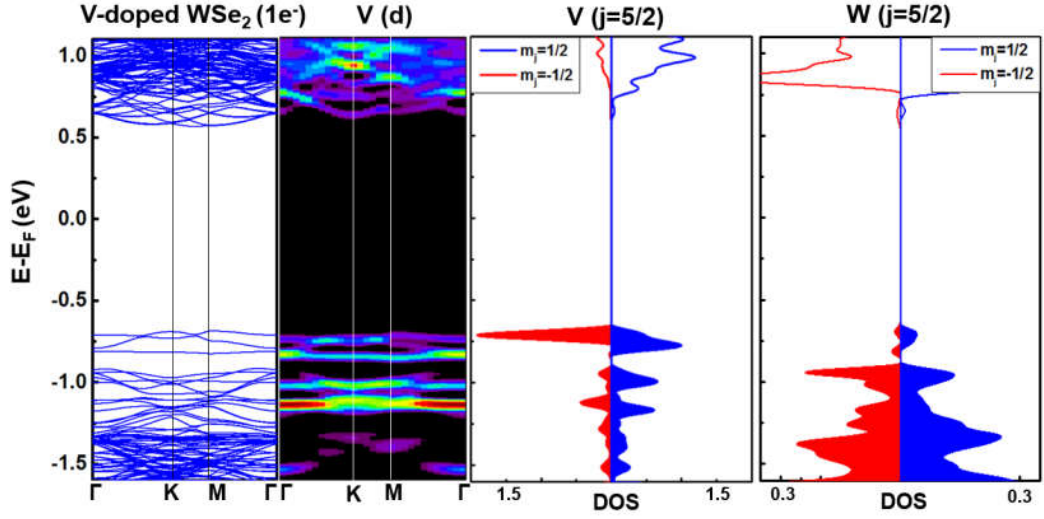

FIG. 4. (a) Band structure of the antiferromagnetic state of V-doped WSe$_2$ at the carrier compensation point (-1e is injected). (b) Local and projected DOSs of (c) V and (d) W atoms.

TABLE 1. Magnetic moments accumulated at the V and W atoms with different carrier densities. $W_n$ and $Se_n$ are the W and Se atoms far from the V site with different distances. Charge – (+) refers to the electron (hole).

|  |  | V | Se$_1$ | Se$_2$ | Se$_3$ | Se$_4$ | W$_1$ | W$_2$ | W$_3$ | W$_4$ |
|---|---|---|---|---|---|---|---|---|---|---|
| Distance from V site (Å) |  | 0 | 2.47 | 4.13 | 7.02 | 10.14 | 3.28 | 6.55 | 9.83 | 13.11 |
| Magnetic moments ($\mu_B$) | charge -1 | 0.459 | -0.020 | 0.000 | 0.000 | 0.000 | -0.032 | 0.000 | 0.000 | 0.000 |
|  | charge 0 | 1.077 | -0.034 | 0.006 | 0.000 | 0.000 | -0.017 | 0.007 | 0.004 | 0.003 |
|  | charge +1 | 1.175 | -0.035 | 0.006 | 0.001 | 0.000 | -0.002 | 0.009 | 0.005 | 0.004 |
|  | charge +2 | 1.189 | -0.035 | 0.006 | 0.001 | 0.000 | 0.003 | 0.007 | 0.002 | 0.001 |